\documentstyle[aps,prb,psfig]{revtex} 
\def\fm{\phi_{\mbox{\scriptsize max}}}

\def\ra{\rightarrow}

\font\tenbb=bbold10 scaled \magstep1

\font\sevenbb=bbold7 scaled \magstep1

\font\fivebb=bbold5 scaled \magstep1

\newfam\bbfam
\textfont\bbfam=\tenbb
\scriptfont\bbfam=\sevenbb
\scriptscriptfont\bbfam=\fivebb
\def\bb{\fam\bbfam\tenbb}

\def\Rb{\mbox{$\bb R$}}

\def\1b{\mbox{$\bb 1$}}

\def\be{\begin{equation}}
\def\ee{\end{equation}}
\def\ra{\rightarrow}

\begin{document}

\title{Arctic octahedron in three-dimensional rhombus tilings
and related integer solid partitions}

\date{\today}

\author{M. Widom$^1$, R. Mosseri$^2$, N. Destainville$^{3,\dagger}$, 
F. Bailly$^4$}
\address{
$^1$Carnegie Mellon U. Department of Physics, Pittsburgh, PA  15213, USA.\\
$^2$Groupe de Physique des Solides, 
Universit\'es Paris 6 et 7, 2 place Jussieu, 
75251 Paris Cedex 05, France.\\
$^3$Laboratoire de Physique Quantique, UMR CNRS-UPS 5626,
Universit\'e Paul Sabatier,
31062 Toulouse Cedex 04, France.\\
$^4$Laboratoire de Physique du Solide-CNRS, 
92195 Meudon Cedex, France. \\
$^{\dagger}$Corresponding author; e-mail: 
{\tt Nicolas.Destainville@irsamc.ups-tlse.fr}
}

\maketitle

\begin{abstract}
Three-dimensional integer partitions provide a convenient
representation of codimension-one three-dimensional random rhombus
tilings. Calculating the entropy for such a model is a notoriously
difficult problem.  We apply transition matrix Monte Carlo simulations
to evaluate their entropy with high precision. We consider both free-
and fixed-boundary tilings. Our results suggest that the ratio of
free- and fixed-boundary entropies is
$\sigma_{free}/\sigma_{fixed}=3/2$, and can be interpreted as the
ratio of the volumes of two simple, nested, polyhedra. This finding
supports a conjecture by Linde, Moore and Nordahl concerning the
``arctic octahedron phenomenon'' in three-dimensional random tilings.

\medskip


\end{abstract}

\section{Introduction}
\label{intro}

Since the discovery of quasicrystals in 1984~\cite{shechtman},
quasiperiodic tilings (such as Penrose tilings) and random rhombus
tilings~\cite{elser85,henley91} have been extensively studied as
paradigms of quasicrystal structure. Quasicrystals are metallic alloys
exhibiting exotic symmetries that are forbidden by usual
crystallographic rules: octagonal, decagonal, dodecagonal or
icosahedral symmetries. Nonetheless, the existence of sharp Bragg
peaks in their diffraction patterns demonstrates a long-range
translational and orientational order.  As compared to {\em perfect}
quasiperiodic tilings, specific degrees of freedom called ``phason
flips'' are active in {\em random} tilings. Despite their random
character, the latter still display the required long-range
quasiperiodic structure.

When tiles are appropriately decorated with atoms, random tilings
become excellent candidates for modeling real quasicrystalline
materials~\cite{AlNiCo}. Therefore the statistical mechanics of random
tilings is of fundamental interest for quasicrystal science. In
particular, the configurational entropy of their phason fluctuations
contributes to the free energy of quasicrystal, and might be a key
ingredient in order to understand quasicrystal stability~\cite{WSS}.

In parallel, random tilings have become an active field of research in
discrete mathematics and computer science (see
~\cite{proppandco,randall95} for instance), and many challenging
questions remain open for investigation in this field.

The relation between random tilings and integer partitions provides an
important tool for the calculation of random tilings
entropy~\cite{elser84,mosseri93,mosseri93B,destain97,destain01}.
Integer partitions are arrays of integers, together with suitable
inequalities between these integers.  One-to-one correspondences can
be established between integer partitions and tilings of rhombi
filling specified polyhedra. However, it seems that such strictly
controlled ``fixed'' boundary conditions inflict a non-trivial
macroscopic effect on random tilings~\cite{elser84,grensing}, even in the
thermodynamic limit, lowering the entropy per tile below the entropy
with free or periodic boundary conditions. This effect {\em a priori}
makes difficult a calculation of free-boundary entropies {\em via} the
partition method.

This boundary sensitivity is well described, for the simple case of
hexagonal tilings~\cite{proppandco,destain98}, in terms of a
spectacular effect known as the ``arctic circle phenomenon'': the
constraint imposed by the boundary effectively freezes macroscopic
regions near the boundary, where the tiling is periodic and has a
vanishing entropy density. Outside these ``frozen'' regions the
entropy density is finite and we call the tiling ``unfrozen''.  The
boundary of the unfrozen region appears to be a perfect circle
inscribed in the hexagonal boundary. The entropy density varies
smoothly within the unfrozen region, reaching a maximum equal to the
free boundary entropy density at the center.

This quantitative result has never been generalized to higher
dimension or codimension tilings, because its generalization requires
the knowledge of the free boundary entropy if one wants to use the
same proof as in the hexagonal case. However, thanks to numerical
simulations, it has recently been conjectured by Linde, Moore,
Nordhal~\cite{moore01} that in dimensions higher than 2, the
corresponding arctic region should be a polytope itself, with flat
boundaries. It was further conjectured (by Destainville and
Mosseri~\cite{DM01} and independently by Propp~\cite{Propp}) that in
this case the entropy density should be spatially uniform and maximal
in the unfrozen region. These conjectures renew the interest for the
partition method since the relation between both entropies becomes
amazingly simple in this case.

On the other hand, except an early Ansatz~\cite{mosseri93} and some
exact numerical results for small tilings~\cite{destain97} (up to
about 300 tiles), almost nothing is known about the entropy of
codimension-one tilings of dimension larger than 2. Note also that
some numerical Monte Carlo simulations provided an estimate of the
entropy of (codimension 3) 3-dimensional random tilings with icosahedral
symmetry~\cite{Strandburg91}.

The present paper is devoted to a numerical investigation of
codimension-one three-dimensional tilings. Thanks to a powerful
transition matrix Monte Carlo algorithm, we achieve precise estimates
of both fixed- and free-boundary entropies. The latter is calculated
via a modified partition method, which produces tilings with fixed
boundaries that do not impose any strain to the tilings, thus
generalizing a former two-dimensional approach~\cite{destain98}.
Comparing both entropies, we support the above conjecture with good
confidence.

The paper is organized as follows: section~\ref{ptbc} reviews the
relation between random tilings and integer partitions, introduces the
different boundary conditions considered in this paper, and describes
the arctic region phenomenon conjecture.  Section~\ref{mc} describes our
Monte Carlo method and our numerical results. Discussions and
conclusions are displayed in the last section.

\section{Partitions, tilings and boundary conditions}
\label{ptbc}

\subsection{Generalities}
\label{gene}

In this paper we consider three-dimensional tilings of rhombohedra
which tile a region of Euclidean space without gaps or overlaps.  A
standard method~\cite{elser85,cutandproject} to generate tilings of
rhombohedra (or of rhombi in two dimensions) consists of selecting
sites and tiles in a $D$-dimensional cubic lattice, then projecting
them into a $d$-dimensional subspace with $D>d$. The difference $D-d$
is called the codimension of the tilings.  The class of symmetry of a
tiling is determined by both its dimension and its codimension and we
denote tilings with such a symmetry by $D \ra d$ tilings.  We consider
in this paper codimension-one, three-dimensional random tilings
(i.e. $D=4$ and $d=3$). These $4 \ra 3$ tilings are composed of four
different rhombohedra. All rhombohedra have identical shapes but they
occur in four possible different orientations. Each rhombohedron is
the projections of one of the four different three-dimensional faces
of a four-dimensional hypercube. The interested reader can refer to
[\cite{destain97}] for a review on codimension-one tilings.

The correspondence between codimension-one tilings and {\em solid
partitions}, a reformulation of the cut-and-project method, is
analyzed in detail in reference [\cite{destain97}], and generalized to
higher codimensions in reference [\cite{destain01}].

We now define three-dimensional solid partitions.  Consider a
three-dimensional array of sides $k_1 \times k_2 \times k_3$.  Fix an
integer $p>0$, called the {\em height} of the partition problem. Put
non-negative integers in the array, no larger than $p$, with the
constraint that these integers decrease in each of the three
directions of space. More precisely, if $i_1$, $i_2$ and $i_3$ are
indices attached to the boxes of the array ($1 \leq i_\alpha
\leq k_\alpha$), we denote by $n_{i_1,i_2,i_3}$ the integral variables
attached to these boxes (the {\em parts}). Our partition array
contains $N_p=k_1 k_2 k_3$ parts. The partition constraint is
\begin{equation}
0 \leq n_{i_1,i_2,i_3} \leq p
\end{equation}
and
\begin{equation}
n_{i_1,i_2,i_3} \geq n_{j_1,j_2,j_3}
\end{equation}
whenever $i_1 \leq j_1$, $i_2 \leq j_2$ and $i_3 \leq j_3$. 

\begin{figure}[ht]
\begin{center}
\ \psfig{figure=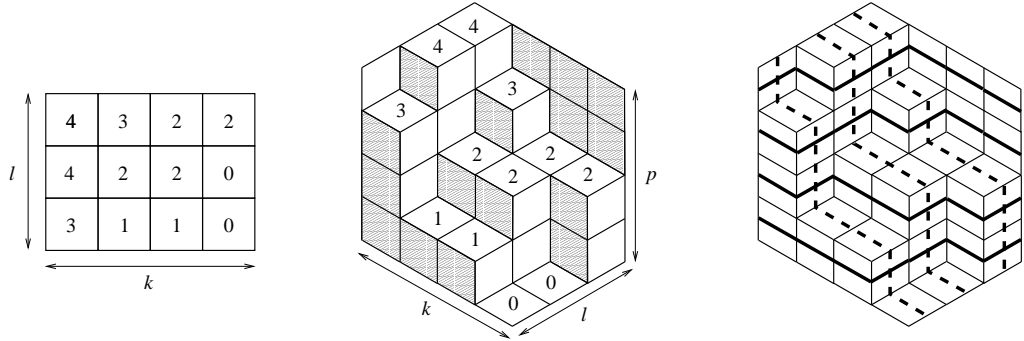,height=4.5cm} \
\end{center}
\caption{One-to-one correspondence
between {\em plane} partitions (left), {\em hexagonal} tilings
(center), and de Bruijn lines (right) in two
dimensions~\protect{\cite{mosseri93,destain97}}; De Bruijn lines of
two families (among three) have been represented.  
\label{mapping32}}
\end{figure}

Figure~\ref{mapping32} displays the tiling/partition correspondence in
two dimensions for easy visualization. Plane partitions are $k \times
l$ arrays of non-negative integers smaller than a {\em height} $p$ and
decreasing in each row and column (left).  This correspondence is
constructed as follows: stack 3-dimensional cubes above the boxes of
the partition so that the height of each stack (center) equals the
value of the corresponding partition box (left). Then project this
stacking along the (1,1,1) direction of the cubic lattice. Faces of
the cubes project to rhombi. The so-obtained rhombus tiling fills a
hexagon of sides $k$, $l$ and $p$ (center).

Following the same construction in three dimensions, four-dimensional
hypercubes are stacked above the three-dimensional partition array,
with the heights of the stacks equal to the corresponding parts. Then
project into three dimensions along the (1,1,1,1) direction of the
hypercubic lattice. Like in the hexagonal case, the so-obtained
tilings fill a polyhedron, a ``rhombic dodecahedron'' ($RD$) of
integral sides $k_1$, $k_2$, $k_3$ and $p$ (see the outer frame in
figure~\ref{octa}). The total number of tiles,
\begin{equation}
N_t=k_1 k_2 k_3 + k_1 k_2 p + k_1 k_3 p + k_2 k_3 p.
\label{NtRD}
\end{equation}
We call tilings with rhombic dodecahedron
boundaries ``$RDB$-tilings'' and denote their configurational entropy
per tile~\cite{footnotea} by $\sigma_{fixed}$.

\begin{figure}[ht]
\begin{center}
\begin{tabular}{ccc}
\parbox{3in}{\vfill \psfig{figure=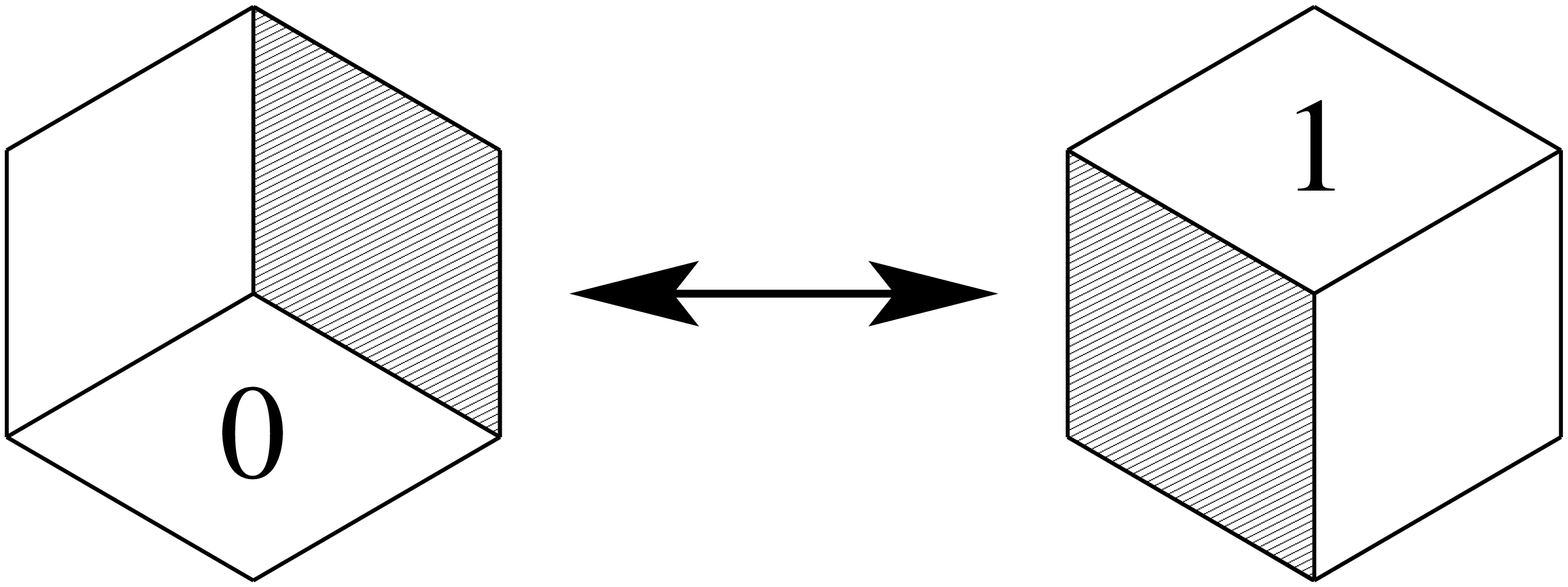,width=3.7cm} \vfill} & \phantom{AAAA} &
\parbox{3in}{\vfill \psfig{figure=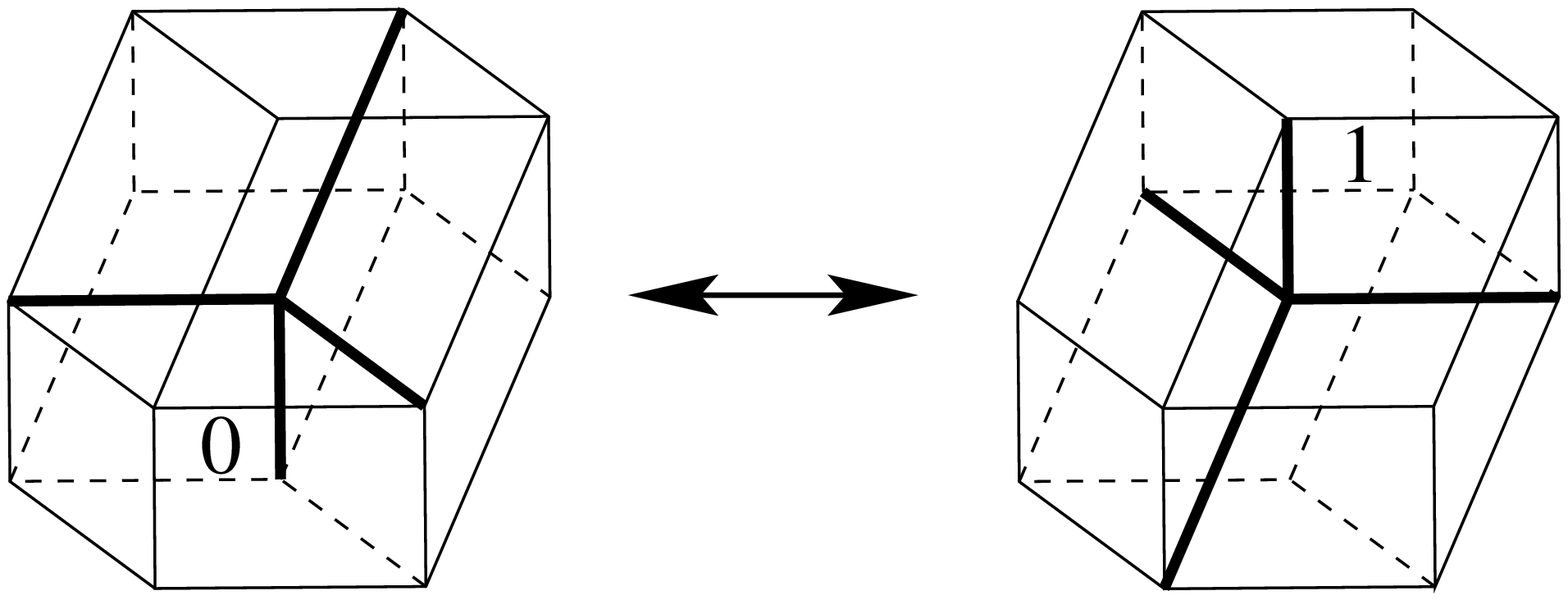,width=6cm} \vfill} \\
(a) & &(b) \\
\end{tabular}
\end{center}
\caption{Examples of single vertex flips: (a) rotation of 3 rhombi inside 
hexagon in $3 \ra 2$ tiling; (b) rotation of 4 rhombohedra inside
rhombic dodecahedron in $4 \ra 3$ tiling. Each rotation increases or
decreases partition height by 1 unit, as shown.}
\label{flip}
\end{figure}

The source of configurational entropy can be easily understood in
either the tiling or the partition representation. Figure~\ref{flip}
illustrates the basic ``tile flip'' in both the $3 \ra 2$ and the $4
\ra 3$ tilings. In each case interior tiles are rearranged without
disturbing the surface of a region. In terms of the equivalent
partition, the height of one partition element increases or decreases
by one unit. This elementary local move is ergodic, and every legal
tiling can be reached by a succession of such flips.

An alternative description of the same tilings and partitions is in
terms of de Bruijn lines~\cite{debruijn} and
membranes. Figure~\ref{mapping32} (right) shows the de Bruijn lines of
the tiling (center). These are formed by connecting centers of
parallel edges on every rhombus. There are three families of de Bruijn
lines, each family with an average orientation perpendicular to the
tile edges. Although de Bruijn lines within a family never cross, de
Bruijn line crossings among different families occur at the center of
each rhombus. For three-dimensional tilings, we have instead de Bruijn
membranes. Elementary flips such as those illustrated in
Fig.~\ref{flip} bend the membrane locally, and membrane fluctuations
become the source of configurational entropy.

Finally, we note that every tiling can be represented as a
$d$-dimensional directed hypersurface embedded in a $D$-dimensional
space.  Beginning with $d=2$ tilings, this hypersurface consists of
the faces of the stacking of cubes visible from the (1,1,1) direction
(figure~\ref{mapping32} (center)). When projected along the same
direction on the ``real'' space $E$, it becomes a plane tiling. It is
``directed'' because no tile overlaps occur during the projection. The
same relationship holds for $d=3$ tilings: the directed hypersurface
consists of the three-dimensional faces of the stacking of hypercubes
viewed along the (1,1,1,1) direction. Since the hypersurface is
directed, it can be seen as a single-valued, continuous, piecewise
linear function $\varphi$ from the real space $E=\Rb^3$ to $\Rb$. The
value of $\phi$ is just the height along the (1,1,1,1) axis. In the
thermodynamic limit, these piecewise linear functions $\varphi$ can be
coarse-grained to obtain smooth functions $\phi: \Rb^3 \ra \Rb$ which
only contain large-scale fluctuations of the original corrugated
membranes~\cite{henley91,destain98}.

\subsection{Boundary conditions}
\label{bcond}

Polyhedral boundary conditions, such as the rhombic dodecahedron
bounding $RDB$ tilings, have macroscopic effects on random tilings. In the
``thermodynamic limit'' of large system size, the statistic ensemble
is dominated by tilings which are fully random only inside a finite
fraction of the full volume and are frozen in macroscopic domains. By
frozen, we mean they exhibit simple periodic tilings in these domains
with a vanishing contribution to the entropy. In two dimensions, this
is known as the ``arctic circle phenomenon'', as described in
introduction~\cite{proppandco,destain98}.

Such boundary conditions are not very physical. Indeed, even if one
can imagine situations where the quasicrystal is constrained by a flat
interface ({\em e.g.} growth experiments on a crystalline substrate),
the previous considerations rely on the assumption that, in the
physical quasicrystalline material, the tiles are elementary
unbreakable structures. However, the system could possibly lower its
free energy by breaking some tiles (a line of tiles in 2D or a surface
in 3D), in order to adapt to the constraint, and thereby free the
remainder the tiling from the constraint.  For sufficiently large
tilings a net lowering of free energy results.  For example, the
energy cost of a broken line of tiles grows linearly with the length
of this line, whereas the free energy difference between a constrained
tiling and a free one grows like the number of tiles and therefore
like length squared (in two dimensions). In the thermodynamic limit,
even if it costs a great amount of energy to break a tile, the system
will eventually prefer to pay this cost. Therefore, the ``true''
thermodynamic entropy density is the free-boundary entropy~\cite{caveat}.

Consequently, it is desirable to relate fixed boundary condition
entropies to the more physical free boundary ones. Fortunately, there
exists an exact formal relation between these
entropies~\cite{destain98}.  A quantitative relation has even been
calculated in the hexagonal case~\cite{proppandco,destain98}, but it
has not been possible (so far) to extend quantitative relationships to
more general classes of tilings.  However, a conjecture by Linde,
Moore and Nordahl~\cite{moore01} has recently brought some new hope (see
section~\ref{arctic}).

To exploit the calculational advantages of a partition representation,
while achieving the physical free-boundary entropy in the
thermodynamic limit, we adapt the partition method so that the
corresponding tilings exhibit no frozen regions.  The new boundary,
even though fixed, has no macroscopic effect on tiling entropy in the
thermodynamic limit. The tilings become homogeneous, displaying the
free-boundary entropy density throughout.

\begin{figure}[ht]
\begin{center}
\ \psfig{figure=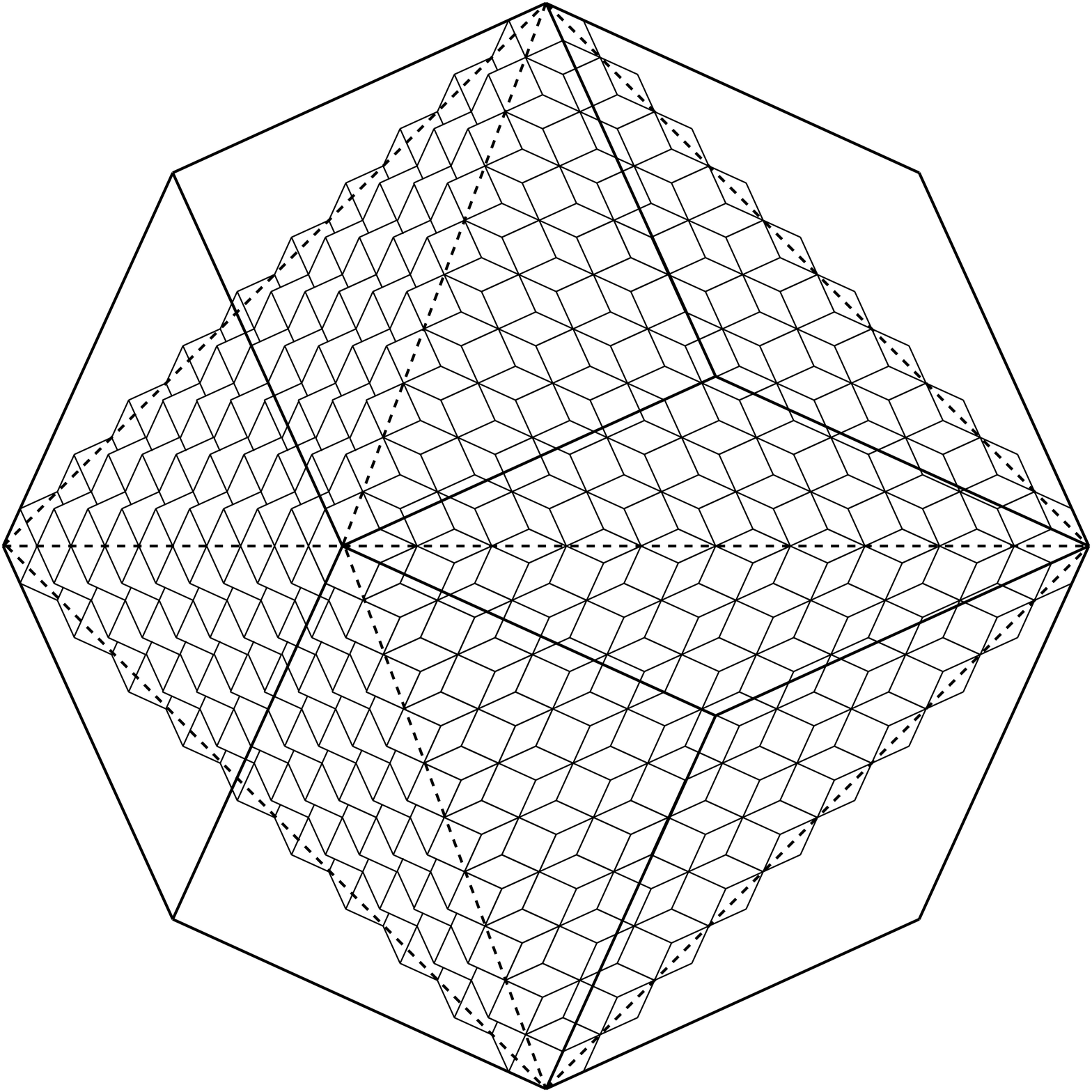,width=7.5cm} \
\end{center}

\caption{The puckered octahedral boundary conditions
($k_1=k_2=k_3=p$). The octahedron $O$ is inscribed in the rhombic
dodecahedron $RD$ of side $p$ coming from the solid partition
method. The volume ratio $|RD|/|O|=3/2$.}
\label{octa}
\end{figure}

The idea is to consider tilings (we focus on ``diagonal'' tilings with
$k_1=k_2=k_3=p$) that fill a regular octahedron $O$ instead of the
rhombic dodecahedron $RD$. Eight vertices of the $RD$ must be
truncated to produce the $O$. We call tilings with octahedron
boundaries $OB$-tilings. Such an octahedron is displayed in
figure~\ref{octa}. It is inscribed in an $RD$ and has puckered
boundaries instead of flat ones. Despite this puckering, the
boundaries are effectively flat in the thermodynamic limit.  The same
kind of idea is developed in reference~\cite{destain98} in the
hexagonal case where a puckered two-dimensional hexagonal boundary is
introduced so that the tilings are homogeneous and exhibit no frozen
regions. It is demonstrated (and numerically checked) that the random
tilings filling this puckered hexagon display a free-boundary entropy
in the thermodynamic limit.

We now explain why $OB$-tilings have a free boundary entropy even
though their boundary is fixed.  The tiling entropy is contained in
fluctuations of the associated hypersurface $\varphi$.  Small-scale
fluctuations are integrated in an entropy functional $s[\phi]$ which
takes into account the total number of possible piecewise linear
hypersurfaces $\varphi$ that are close to the smooth one $\phi$ (see
reference [\cite{destain98}] for discussion). Functions maximizing
$s[\phi]$ represent the dominant macroscopic states of the
system. Note that $s[\phi]$ can be written as a functional of the
gradients of $\phi$, known as the {\em phason} gradient or {\em phason
strain} in the quasicrystal community. And $s[\phi]$ is maximum and
equal to the free-boundary entropy $\sigma_{free}$ when this gradient
vanishes everywhere.

Fixed boundaries on tilings translate into fixed boundary conditions
for the functions $\phi$. Therefore $s[\phi]$ must be maximized on a
restricted set of functions, $F$. For rhombic dodecahedral boundaries
on $RDB$ tilings, the boundaries of functions $\phi \in F$ are non-flat
polyhedra, and the phason gradient cannot vanish everywhere. Therefore
their entropy density $\sigma_{fixed}$ is bounded below by
$\sigma_{free}$. For octahedral boundaries on $OB$ tilings, the
functions $\phi$ are also constrained by a fixed boundary
condition. But in this case, the boundary is flat and {\em
strain-free}. It does not impose any phason gradient on the functions
$\phi$. The phason gradient {\em can} vanish everywhere and the maximum of
$s[\phi]$ equals $\sigma_{free}$.

\begin{figure}[ht]

\begin{center}
\ \psfig{figure=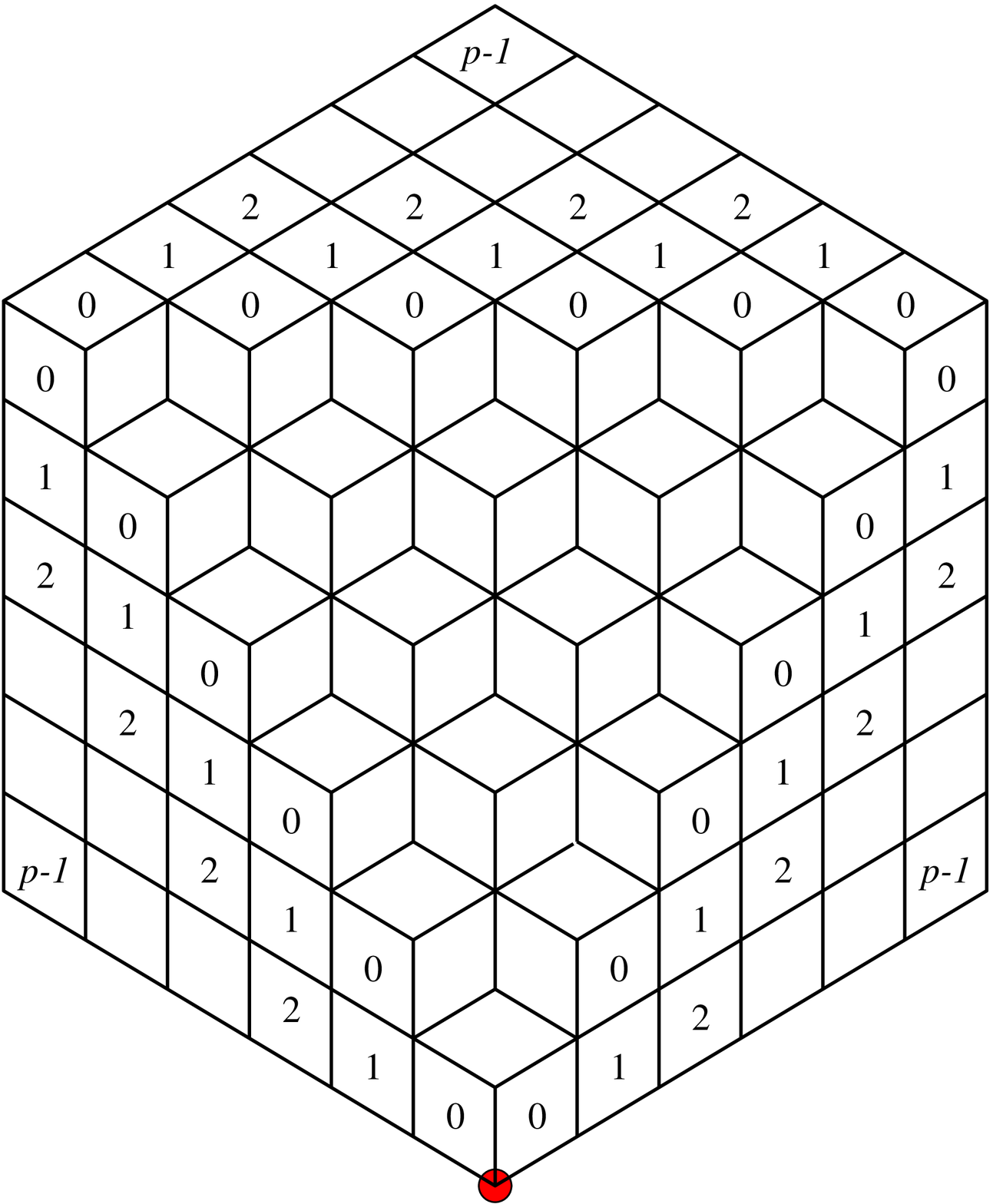,width=5cm} \

\ \psfig{figure=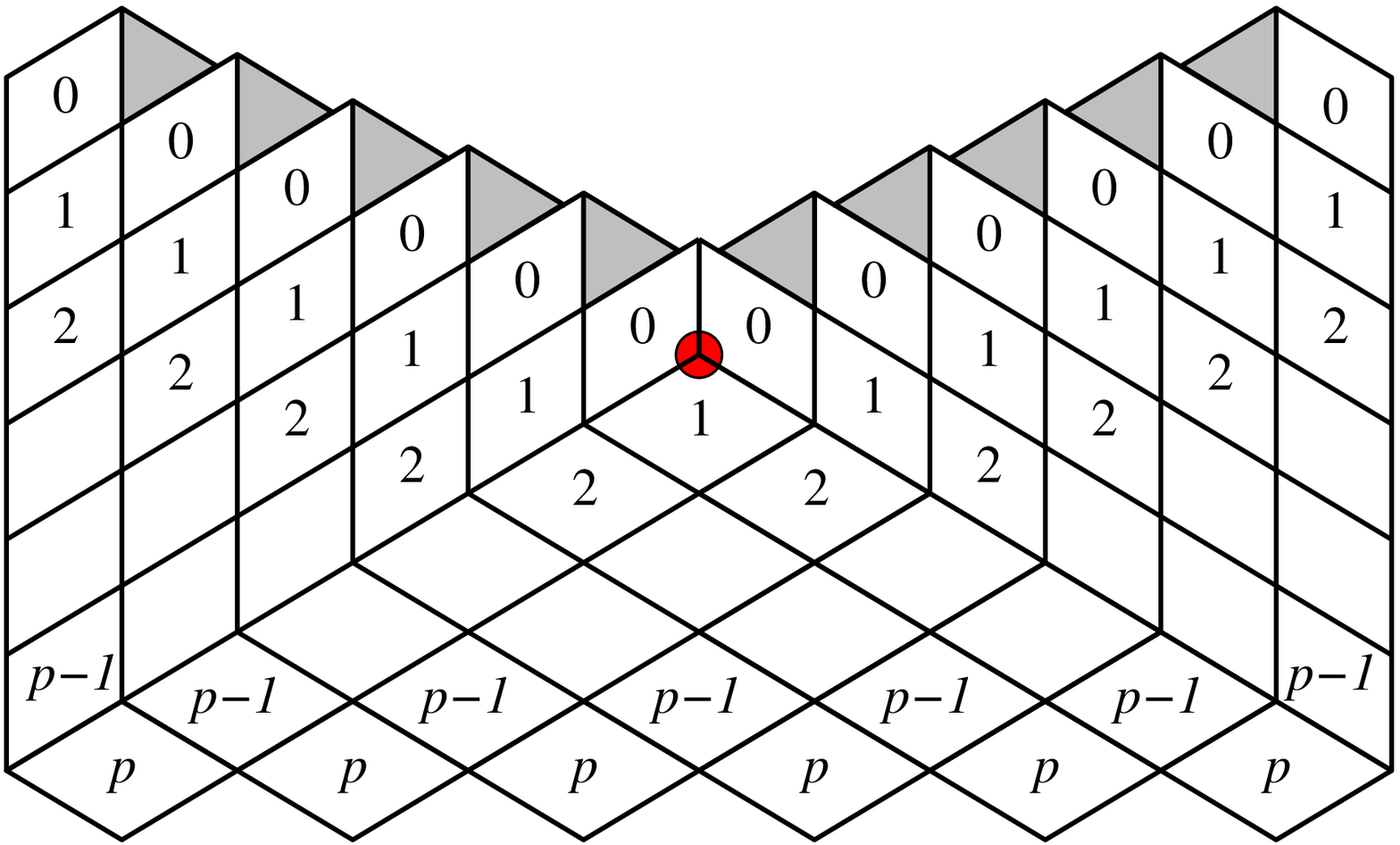,width=5cm} \
\end{center}

\caption{Boundary conditions for the partition problem associated with
the strain-free tiling problem with boundary conditions of
Fig.~\protect{\ref{octa}}. Two views of the truncated cube (of side
$p$) are provided. The marked vertex is the same in both views.}
\label{bords}
\end{figure}

Figure~\ref{bords} illustrates changes in the partition array boundary
conditions needed to achieve $OB$ tilings instead of $RDB$ tilings.  The
partition array is no longer a cube, but two opposite pyramidal
corners have been truncated, leaving a slab of $D_{3d}$ symmetry. This
slab contains $N_p=p^3-(p-1)p(p+1)/3$ parts.

We indicate minimal and maximal values of the parts: the three faces in
the topmost figure bear minimal values of their adjacent boxes,
whereas the three hidden faces bear maximal values. These values are
visible on the bottom in the view in the bottommost figure.
Individual parts are bounded by integers depending on the position of
the part in the array. Since interior parts are bounded by their
neighbors, it is sufficient to impose these bounds on surface parts,
as indicated in figure~\ref{bords}. The lower bounds range from 0 to
$p-1$, with constant values on lines parallel to diagonal of cube
faces. The upper bounds are read on the opposite faces, and range from
1 to $p$.

Using the methods discussed in section~\ref{gene}, such partitions
generate tilings filling the octahedron $O$. For an original partition
cube of sides $k_1=k_2=k_3=p$, the full $RD$ contains
$N_t=4p^3$ tiles (see equation~(\ref{NtRD})). The octahedron $O$
contains $N_t = 4p^3 - 4(p-1)p(p+1)/3$ tiles (these are the $4p^3$
tiles in the $RD$ minus the tiles in the truncated corners). These
partitions are used in section~\ref{mc} to estimate numerically
$\sigma_{free}$.

\subsection{Arctic octahedron conjecture (Linde, Moore and Nordahl)}
\label{arctic}

The relation between free and fixed boundary entropies in not known in
general, because the generalization of the proof of the hexagonal case
requires the knowledge of the free boundary entropy. In two
dimensions, this relation is non-trivial because the unique function
$\fm$ maximizing $s[\phi]$ has a complex
expression~\cite{proppandco,destain98}, leading to the arctic circle
phenomenon: the statistically dominating tilings are frozen (and
periodic) outside a circle inscribed in the boundary hexagon.

In three dimensions, Linde, Moore and Nordahl have recently explored
numerically the typical shape of a $RDB$-tiling and have
conjectured~\cite{moore01} that the two-dimensional circle then
becomes not a sphere but a regular octahedron, inscribed in $RD$ like
$O$ in figure~\ref{octa}. More precisely, the unfrozen region is not
exactly $O$ but tends towards $O$ at the large size limit.

This conjecture has a crucial consequence~\cite{DM01}: the statistical
ensemble of $RDB$-tilings is dominated by tilings periodic outside $O$
and random inside $O$, equivalently $OB$-tilings like in the previous
section completed by eight periodically tiled pyramids to fill
$RD$. Note that James Propp~\cite{Propp} has also conjectured
independently of the present work that $RDB$-tilings should be
homogeneous inside $O$.  Taking into account the tiles in the frozen
regions to calculate an entropy per tile, one finally gets
\begin{equation}
\label{ratio}
\sigma_{fixed} = {{N_{O}}\over{N_{RD}}} \sigma_{free}
= {2 \over 3} \sigma_{free},
\end{equation}
since the ratio of the numbers of tiles in $RD$ and $O$ is 3/2. The
next section is dedicated to the calculation of this ratio {\em via}
Monte Carlo simulations.

\section{Monte Carlo}
\label{mc}

Conventional Monte Carlo simulations using Metropolis
sampling~\cite{qcmc} are useful for generating typical tiling
configurations.  They can reveal temperature driven phase
transitions~\cite{PT} and yield quantitative evaluations of phason
elastic constants from fluctuations. Evaluation of entropy by Monte
Carlo simulation demands specialized techniques.  Advanced methods
using specialized dynamics based on entropic sampling~\cite{entropic}
and transfer matrices~\cite{qctmmc,qctm} yield reasonably accurate free
energies and absolute entropies.

We employ a variant of the transition matrix
method~\cite{TMMC,footnoteb} that couples a conventional Metropolis
Monte Carlo simulation with a novel data collection and analysis
scheme to construct a numerical approximation to the {\em
transition matrix}, described below in section~\ref{tm}. The
density of states is an eigenvector of the exact transition matrix,
and the sum of the density of states yields the total number of
states. This method yields highly accurate absolute entropies with
impressive efficiency.

\subsection{Transition Matrix}
\label{tm}

For any legal partition $P=\{n_{ijk}\}$, we define its ``energy'' as
its total height
\be
\label{Ham}
E(P)=\sum_{ijk} n_{ijk},
\ee
and incorporate the constraint of legality by defining $E=\infty$ for
any partition $P$ that violates the ordering conditions inside the
partition or at the boundaries. Under single vertex flip dynamics (see
Fig.~\ref{flip}) a single partition element increases or decreases by
one unit resulting in an energy change $\Delta E = \pm 1$.  The ground
state of this model is the lowest height legal partition. For the
boundary conditions employed here, the ground state is unique and we
denote its energy as $E_{min}$. There is also a unique {\em maximum}
energy state of energy $E_{max}$.

In a Metropolis Monte Carlo simulation of the partition, a randomly
chosen partition element is randomly increased or decreased by
one. The change is accepted if it lowers the energy (i.e. we attempted
to decrease it, and the result is a legal partition). If the change
raises the energy, it is accepted with probability
$\exp{(-E/T)}$. Note that illegal partitions are never accepted. Low
temperatures bias the partitions towards low heights, while high
temperatures remove any bias according to height. As $T$ becomes large,
the Metropolis simulation faithfully reproduces the random tiling
ensemble in which all configurations have equal weight. If we choose,
we may take negative values of temperature $T$. Large negative
temperatures again reproduce the random tiling ensemble, while small
negative temperatures favor partitions of maximal height.

For a given partition $P$, a number of partitions $n_{\pm}(P)$ can be
reached by single upwards or downwards steps. Additionally, a certain
number $n_0(P)$ of steps are forbidden due to the partition constraints.
The sum rule
\be
\label{sumrulen}
n_{-}(P)+n_{0}(P)+n_{+}(P)=2N_p
\ee
holds for every partition $P$. This value is twice the number of parts
because each part can be raised or lowered. For any given partition
$P$, the calculation of $n_{\pm,0}$ is easy (though not fast) and
exact.

We define the {\em transition matrix} $\omega_{\pm,0}(E)$ as a matrix
with $E_{max}-E_{min}+1$ rows (one for each allowed energy $E$) and
three columns (labeled ``+'' for upwards transitions, 0 for forbidden
transitions and ``-'' for downwards transitions. Alternatively, we can
think of these three columns as the diagonal and two off-diagonals of
a square matrix $\omega(E,E')$ of dimension $E_{max}-E_{min}+1$.
Formally, we define
\be
\label{rates}
\omega_{\pm,0}(E) \equiv {{1}\over{W(E)}} \sum_{P(E)} n_{\pm,0}(P)/2N_p
\ee
where the sum is over all partitions of energy $E$ and the
normalization $W(E)$ is the total number of partitions with energy
$E$. In general the set $P(E)$ and the value $W(E)$ are not known,
preventing us from actually calculating the transition
matrix. However, we may calculate the matrix numerically with high
precision by averaging $n_{\pm,0}(P)/2N_p$ over those partitions $P$
occurring during a Monte Carlo simulation.  By virtue of
eq.~(\ref{sumrulen}), the matrix elements obey the sum rule
\be
\label{sumrulew}
\omega_{-}(E)+\omega_{0}(E)+\omega_{+}(E)=1
\ee
for any energy $E$, so each row of the transition matrix can be
normalized independently without knowledge of $W(E)$.

The transition matrix can be interpreted in terms of the rates at
which Monte Carlo moves {\em would} be accepted in a {\em
hypothetical} simulation at {\em infinite} temperature (even if our
actual simulation is performed at finite temperature).  In a single
Monte Carlo step, the probability for upwards or downwards transitions
at infinite temperature is $n_{\pm}(P)/2N_p$, and the probability a
move will be rejected is $n_0(P)/2N_p$.  We can predict {\em finite}
temperature transition probabilities by multiplying
$\omega_{\pm,0}(E)$ with appropriate Boltzmann factors, and in fact
these agree well with actual observed acceptance rates.

To extract the density of states and the entropy, consider the
infinite temperature detailed balance condition.  At infinite
temperature the probability a randomly chosen partition has energy $E$
is $W(E)/Z$, where
\begin{equation}
Z=\sum_E W(E)
\end{equation}
is the total number of legal partitions (the ``partition
function''). Multiplying the probability $W(E)/Z$ by the acceptance
rate of upwards transitions $\omega_{+}(E)$ yields the total forwards
transition rate. The backwards rate is obtained in similar
fashion. Detailed balance requires that the total rate of forward
transitions from energy $E$ to energy $E+1$ must equal the total rate
of backwards transitions, hence
\be
\label{balance}
\omega_{+}(E) W(E)/Z=\omega_{-}(E+1) W(E+1)/Z.
\ee
It is useful to rearrange the detailed balance eq.~(\ref{balance}) to
find
\be
\label{iterate}
W(E+1)={{\omega_{-}(E+1)}\over{\omega_{+}(E)}} W(E)
\ee
which allows us to iteratively extract the full density of states
function $W(E)$ using uniqueness of the ground state, $W(E_{min})=1$.
Finally, the total entropy
\be
S=\ln{Z},
\ee
and the entropy density $\sigma=S/N_t$.

Note that these {\em infinite} temperature transition rates may be
calculated from {\em finite} temperature Metropolis
simulations. Indeed, certain low energies $E$ may occur very rarely in
high temperature simulations so we must perform low temperature
simulations to generate partitions with energy $E$, from which the
infinite temperature transition rates may be calculated.  By sweeping
over a range of temperatures we obtain accurate values of
$\omega_{\pm}(E)$ for all energies. Then, we use the infinite
temperature detailed balance condition (\ref{balance}) to extract the
density of states even though our simulation is performed entirely at
finite temperatures.

The accuracy of our result is controlled by the accuracy with which we
determine $\omega_{\pm}(E)$. For each partition, the transition
numbers $n_{\pm,0}(P)$ are calculated exactly and added into the row
of the transition matrix with energy $E(P)$. Since there may be many
partitions with the same energy $E$, each having different transition
numbers, the accuracy with which a row of the transition matrix is
determined is limited by our ability to generate a representative
sample of partitions. We store the matrix as the integer valued sum of
the integers $n_{\pm,0}(P)$, and impose the
normalization~(\ref{sumrulew}) after data collection is
complete. Because the values of $n_{\pm,0}(P)$ can be quite large, and
we visit each energy a very large number of times, ordinary 4 byte
integers can not hold the data. We implemented special procedures to
handle storage and algebraic manipulations of large integers.

\subsection{Numerical Data}
\label{data}

Fixed boundary partitions are initialized at zero height (and thus
$E=E_{min}=0$). Their maximum energy $E_{max}=k_1 k_2 k_3 p$. We
accumulate data in the transition matrix through four sweeps over
temperature: from $T_{min}$ to $T_{max}$ during which time the mean
energy grows from $E_{min}$ to about $E_{max}/2$; from $-T_{max}$ to
$-T_{min}$ during which time the mean energy grows from $E_{max}/2$ to
$E_{max}$; from $-T_{min}$ back to $-T_{max}$; from $T_{max}$ back
down to $T_{min}$. Each sweep visits 201 temperatures in a geometric
sequence from the initial to the final temperature.  Free boundary
tilings are initialized with a flat partition at energy close to
$(E_{min}+E_{max})/2$. We again perform four sweeps: from $T_{max}$
down to $T_{min}$; from $T_{min}$ back up to $T_{max}$; from
$-T_{max}$ to $-T_{min}$; from $-T_{min}$ back to $-T_{max}$.

For both fixed and free boundary tilings, symmetries of the model
dictate that the density of states is symmetric about the midpoint
energy $E_{mid}=(E_{min}+E_{max})/2$. The density of states $W(E)$ has
a strong maximum at the midpoint.  By including both positive and
negative temperature sweeps we sample both low and high energy
states. By reversing our sweeps we mitigate possible systematic
sampling errors associated with the direction of the sweep.

At each temperature during a sweep, we accumulate data on $N_{MC}$
sample configurations.  For a given configuration (partition $P$), the
time required to calculate $n_{\pm,0}(P)$ is proportional to the
partition size $N_p$, and hence to the time required to attempt $N_p$
flips.  Because the data accumulation is time consuming, we can
perform many single vertex flips between sample without significantly
slowing down the simulation. We take $N_{FL}=N_p/5$ single vertex
flips, which yields rough equality between time spent flipping and
collecting data. Prior to data collection at any temperature, we
anneal at fixed temperature for $N_{MC}\times N_{FL}/100$ steps. The
total number of attempted flips in an $N_{MC}=10^6$ run on a $10\times
10\times 10$ partition is thus in excess of $4\times 201\times
10^6\times 10^3/5=1.6 \times 10^{11}$ attempted flips. A run of this
length takes 5 days on a 1.7GHz Pentium 4 processor.

\begin{figure}[htb]
\begin{center}
\ \psfig{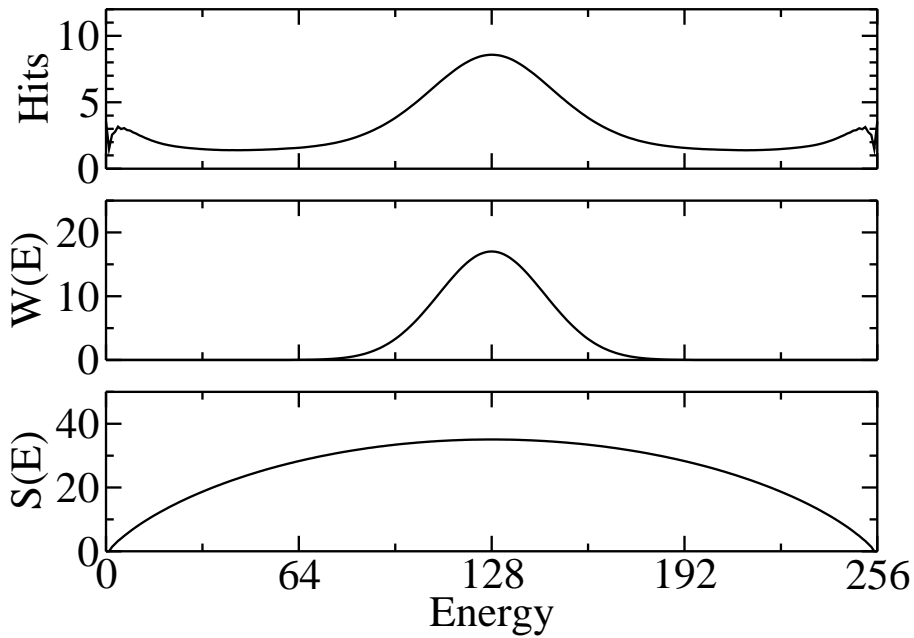} \
\end{center}
\caption{
Simulation data for $p=4$ fixed boundary tiling.  Top panel: Number of
configurations sampled (units of $10^6$). Center panel: Density of
states $W(E)$ (units of $10^{14}$ per energy).  Bottom panel:
Microcanonical entropy $S(E)=\ln{W(E)}$.}
\label{fig:data}
\end{figure}

This protocol was chosen to ensure approximately uniform coverage of
energies from $E_{min}$ to $E_{max}$. The energy distribution of
partitions visited at each temperature overlaps strongly the energy
distribution of partitions visited at the previous and following
temperatures. The value of $T_{min}$ is chosen sufficiently low to
guarantee some coverage of the extreme energy states. Because there
are few states at the energy extremes, there is an entropic barrier to
reaching these extremes. This causes a spike in the coverage close to
the extreme energies that cannot be avoided using Metropolis sampling.
The value of $T_{max}$ is chosen sufficiently high that the midpoint
energy is a local maximum in the coverage. Fig.~\ref{fig:data} plots the
number of partitions sampled as a function of energy during the
longest runs (length $10^6$) for the $p=4$ partition. Note that the
number of hits exceeds $10^6$ uniformly for each energy in the range
$0 \le E \le 256$.

The density of states, $W(E)$ shown in figure~\ref{fig:data} (center),
is nearly a Gaussian. $W(E)$ reaches a peak value of $1.7\times
10^{15}$ states at energy $E=128$, and has a half width at half
maximum of $\Delta E=20$. As the system size grows this width grows
more slowly than the number of tiles, so the density of states
asymptotically approaches a delta function. Although we sampled a
total of $8 \times 10^6$ configurations at $E=128$, this represents a
fraction of only about $4.7 \times 10^{-9}$ of the total number that
exist at this energy.

The microcanonical entropy, defined as $\log{W(E)}$, is plotted in the
lower panel of Fig.~\ref{fig:data}. This plot reveals the expected
symmetry around $E_{mid}$. The degree to which symmetry is broken can
be used as an indicator of errors accumulated during our iterative
calculation~(\ref{iterate}), because sampling errors at low and high
energies need not cancel. By inspection it can be seen that this error
is quite low, since the entropy returns essentially to zero at high
energy.

\vspace{2cm}

Table~\ref{converge} shows the convergence of entropy data as run
length grows. We show the convergence only for the largest system
size, which represents our worst case. The residual
$R=\log{W(E_{max})}$ measures the failure of the calculated
microcanonical entropy to return to zero at high energy. It reflects
the cumulative error in $W(E)$, and thus, assuming statistical
independence of the errors at each energy, it provides an upper bound
on the error in $W(E)$ for any energy. Converting the error in $W(E)$ into
an error in the entropy, we estimate an uncertainty
\begin{equation}
|\Delta \sigma| \approx R/N_t.
\end{equation}
Our numerical simulation results in Table~\ref{converge}, and also
simulations of smaller systems for which the entropy is known exactly,
are in good numerical agreement with this estimate.

Table~\ref{sigmas} summarizes our data for all system sizes
studied. We report the entropy resulting from the run of length
$N_{MC}=10^6$, and use the difference between that run and the length
$N_{MC}=10^5$ run for the quoted uncertainty.

To extrapolate our data to infinity, we fit to functional forms. For
the fixed boundary entropy, we expect finite-size corrections of order
$\log(p)/p$ and $1/p$ because of boundary effects. Indeed, these two
terms are the first correction terms in the exactly solvable one- and
two-dimensional random tilings. Note that they have the same order of
magnitude for the values of $p$ considered here. This logarithmic form
fits the data much better than a simple correction of order $1/p$.
For fixed boundary tilings the data fit well to
\begin{equation}
\sigma_{fixed}(p) \simeq 0.145- 0.0049 {\log(p) \over p}+ {0.034 \over p},
\end{equation}
from which we conclude that $\sigma_{fixed}=0.145(3)$. The value is
obtained from a fit excluding the data point with $p=1$, while the
uncertainty estimate comes from excluding instead the data point with
$p=10$. This value is very close to a conjectured
limit~\cite{mosseri93} of 0.139, but given our small uncertainty, we
believe the conjectured value is not exact.

For the free boundary data we find
\begin{equation}
\sigma_{free}(p) \simeq 0.214- 0.052 {\log(p) \over p} - {0.046 \over p},
\end{equation}
from which we conclude that $\sigma_{free}=0.214(2)$. As in the fixed
boundary case, the logarithmic finite size correction fits the data
better than a simple correction of order $1/p$. Should the octahedral
fixed boundary have a trivial short-range effect, it would create only
a $1/p$ leading term (from surface {\em vs} bulk contributions), as
indeed occurs in the corresponding two-dimensional hexagonal
case~\cite{destain98} (see~\ref{bcond}).  Therefore, it appears that
the strain-free octahedral boundary has a nontrivial long-range effect
on the local entropy density, presumably by limiting the fluctations
of the height function within a ``penetration depth'' that depends on
the side length $p$.  A penetration depth growing like $\log(p)$ is
consistent with our leading finite-size correction.

For the ratio, we fit to
\begin{equation}
{\sigma_{free}(p) \over \sigma_{fixed}(p)} \simeq  1.48
 - 0.37 {\log(p) \over p} - {0.54 \over p},
\end{equation}
and we conclude the ratio $\sigma_{free}/\sigma_{fixed}=1.48(3)$. This
value equals 3/2 within our uncertainty.

It would be desirable to achieve higher precision in this ratio in the
future. At present, we are limited to system sizes of $p=10$ or less
because for larger systems the density of states $W(E)$ exceeds
$10^{308}$, the limiting floating point number on our Pentium 4
processor. Either specialized floating point arithmetic or a 64-bit
processor will be required to treat systems of size $p=11$ and above.

\section{Discussion}
\label{discussion}

Returning to the arctic octahedron conjecture, we recall (see
equation~(\ref{ratio})) we expected a ratio
$\sigma_{free}/\sigma_{fixed}=3/2$.  In fact, with few extra
hypotheses, we now demonstrate that our numerical results provide a
strong support to the arctic octahedron conjecture.  Suppose that, as
in the two-dimensional case~\cite{footnotec}, the function $\fm$
maximizing the entropy functional $s[\phi]$ subject to a given
boundary condition is unique~\cite{destain98}. If $\phi_0$ is the
piecewise linear function which vanishes identically in the central
octahedral region $O$ and has maximum strain (and 0 entropy) in the
eight pyramidal regions comprising $RD-O$, then $s[\phi_0] = 2/3 \;
\sigma_{free}$, and thus $s[\phi_0] = \sigma_{fixed}$. Therefore
$\phi_0=\fm$, by uniqueness of $\fm$.

As a conclusion, if $\sigma_{free}/\sigma_{fixed}=3/2$ and $\fm$ is
unique, then the arctic octahedron conjecture is true. Note that this
conjecture can be generalized to higher dimensions, making in
principle possible a calculation of the above ratio for any dimension,
at least as far as codimension one problems are concerned.

In terms of de Bruijn membranes, the above result means that all
de Bruijn membranes are straight (or flat), at least at large scales. 
Indeed if de Bruijn membranes are flat in a $RD$-tiling, the
4 de Bruijn families intersect in the central octahedron $O$, but only
intersect 3 by 3 in the 8 pyramidal corners, leading to 
a frozen tiling in these 8 regions and a strain-free tiling
in $O$.

By comparison, one-dimensional de Bruijn lines in two dimensional
hexagonal tilings are not straight since that would lead to a hexagonal
central arctic region with uniform entropy density. If
$\sigma_{free}^{3 \ra 2}$ denotes the free boundary entropy of
hexagonal tilings, one would get a fixed boundary diagonal entropy per
tile of $3 \; \sigma_{free}^{3 \ra 2}/4 = 0.242$ instead of 0.261
(using similar arguments as above). Fluctuations of the de Bruijn lines
raise the entropy from 0.242 to 0.261. Entropic repulsion of the lines
causes them to bend.

In contrast, the repulsion between de Bruijn membranes is sufficiently
weak that they are not forced away from their flat configuration. This
is possibly related to the fact that the fluctuation of a free
2-dimensional directed membrane in 3-dimensional space is of order
$\sqrt{\log(L)}$ where $L$ is its linear size~\cite{henley91}, whereas
the fluctuation of a free 1-dimensional directed polymer in
2-dimensional space is larger and of order $\sqrt{L}$.  Therefore it
is natural to suppose that the flatness of the arctic region will
persist in dimensions higher than 3 where the fluctuations are even
smaller, since they are bounded~\cite{henley91}.

Our result emphasizes important dimensional dependence of the
transition between the frozen and unfrozen regions. Indeed, in 2
dimensions, the transition is continuous, since the entropy density is
0 by the arctic circle and then continuously varies to reach its
maximum value near the center of the hexagon, with a non-zero gradient
everywhere except near the center. By contrast, the situation seems to
be radically new and different in 3 dimensions, since the entropy
density appears to be constant in the arctic octahedron $O$, with a
vanishing gradient everywhere and a discontinuous transition at the
boundary of $O$.

This result (as well as its possible generalization to higher
dimensions) is a strong support in favor of the partition method, of
which it was formerly believed that it could not easily provide
relevant results about free boundary entropies. Indeed, provided the
arctic region is polyhedral {\em and} its boundary is strain-free, the
ratio of both entropies is nothing but a ratio of volumes of suitable
polytopes. The latter two conditions should be fulfilled as soon as
the entropic repulsion between de Bruijn membranes is sufficiently
weak, that is as soon as the spatial dimension is 3 or greater.

To finish with, we mention that this transition matrix Monte Carlo
technique can easily be adapted to the numerical calculation
of the entropy of two-dimensional rhombus tilings. Indeed, the
structure of the configuration space of such tiling problems
in terms of flips has been characterized in reference~\cite{destain01}.
Note however that no such simple result as the ``arctic octahedron
phenomenon'' is expected in these two-dimensional classes of tilings,
but the calculation of their configurational entropy is a challenge 
in itself. This work is in progress.

\acknowledgments
We acknowledge useful conversations with James Propp and 
Cristopher Moore. We thank Robert Swendsen for suggesting 
the computational approach.  This work was supported in 
part by NSF grants DMR-0111198 and INT-9603372.

\newpage

\begin{table}[htb]
\caption{Convergence of $p=10$ entropy data for increasing run length. The
residual $R=\log{W(E_{max})}$ measures the cumulative error in $W(E)$.}
\label{converge}
\begin{tabular}{||l|r|r|r|r||}
$N_{MC}$         &  $10^3$   &   $10^4$    &  $ 10^5$  &  $10^6$  \\
$\sigma_{fixed}$ &  0.147827 &   0.147385  &  0.147400 &  0.147349   \\
$R_{fxixed}$     &  3.474    &   0.180     &  0.429    &  0.002    \\
$\sigma_{free}$  &  0.197516 &   0.197300  &  0.197234 &  0.197273 \\
$R_{free}$       &  1.214    &   0.143     & -0.202    & -0.036  \\
\end{tabular}
\end{table}

\begin{table}[htb]
\caption{Size-dependent entropies. Values in parentheses are uncertainties
in final digit. Values without uncertainties are exact.}
\label{sigmas}
\begin{tabular}{||r|l|l|l||}
$p$   & $\sigma_{free}$ & $\sigma_{fixed}$ & $\sigma_{free}/\sigma_{fixed}$ \\
\hline
1     & 0.1732868       &  0.1732868       & 1.000 \\
2     & 0.1732868       &  0.1601239       & 1.080 \\
3     & 0.17947(2)      &  0.1545769       & 1.161 \\
4     & 0.18455(6)      &  0.1517949       & 1.216 \\
5     & 0.18829(3)      &  0.15017(2)      & 1.254 \\
6     & 0.19108(4)      &  0.14918(6)      & 1.281 \\
7     & 0.19320(4)      &  0.14848(1)      & 1.301 \\
8     & 0.19486(2)      &  0.14780(1)      & 1.318 \\
9     & 0.19618(1)      &  0.14762(2)      & 1.329 \\
10    & 0.19727(4)      &  0.14735(5)      & 1.339 \\
\hline
$\infty$& 0.214(2)      &  0.145(3)        & 1.48(3) \\
\end{tabular}
\end{table}

\newpage


\begin{thebibliography}{99}
\bibitem{footnotea} Recall that the configurational entropy per tile
is the logarithm of the number of configurations, divided by the
number of tiles.
\bibitem{caveat} To make the ``free-boundary'' entropy well defined, 
we need to add some weak constraint to ensure the tiling remains
connected and reasonably compact. We presume that the free-boundary
entropy equals that for strain-free periodic boundary conditions.
\bibitem{footnotec} It was shown in reference~\cite{destain98} that
the function $\fm$ is unique provided the free-boundary entropy is a
concave function of the phason gradient.
\end{thebibliography}
\end{document}